\documentclass{article}

\usepackage{arxiv}

\usepackage[utf8]{inputenc} % allow utf-8 input
\usepackage[T1]{fontenc}    % use 8-bit T1 fonts
\usepackage{hyperref}       % hyperlinks
\usepackage{url}            % simple URL typesetting
\usepackage{booktabs}       % professional-quality tables
\usepackage{amsfonts}       % blackboard math symbols
\usepackage{nicefrac}       % compact symbols for 1/2, etc.
\usepackage{microtype}      % microtypography
\usepackage{lipsum}
\usepackage{graphicx}
\usepackage{booktabs}
\usepackage{multirow}
\usepackage{graphicx}
\usepackage[table,xcdraw]{xcolor}
\graphicspath{ {./images/} }

\title{Glorbit: A Modular, Web-Based Platform for AI Based Periorbital Measurement in Low-Resource Settings}

\author{
George R. Nahass\thanks{These authors contributed equally.} \\
Department of Biomedical Engineering\\
University of Illinois Chicago\\
Chicago, IL, USA \\
\texttt{gnahas2@uic.edu} \\
\And
Jacob van der Ende\footnotemark[1] \\
Quina Care Hospital\\
San Miguel, Putumayo\\
Ecuador \\
\texttt{jacob.vanderende@quinacare.org} \\
\And
Sasha Hubschman \\
Department of Ophthalmology\\
University of Illinois Chicago\\
Chicago, IL, USA \\
\And
Benjamin Beltran \\
Department of Biomedical Engineering\\
University of Illinois Chicago\\
Chicago, IL, USA \\
\And
Bhavana Kolli \\
Department of Ophthalmology\\
University of Illinois Chicago\\
Chicago, IL, USA \\
\And
Caitlin Berek \\
Department of Ophthalmology\\
University of Illinois Chicago\\
Chicago, IL, USA \\
\And
James D. Edmonds \\
Department of Ophthalmology\\
University of Illinois Chicago\\
Chicago, IL, USA \\
\And
R.V. Paul Chan \\
Department of Ophthalmology\\
University of Illinois Chicago\\
Chicago, IL, USA \\
\And
Pete Setabutr* \\
Department of Ophthalmology\\
University of Illinois Chicago\\
Chicago, IL, USA \\
\And
James W. Larrick \\
Panorama Research Institute\\
Palo Alto, CA, USA\\
Quina Care Hospital\\
San Miguel, Putumayo, Ecuador \\
\And
Darvin Yi \\
Department of Biomedical Engineering\\
Department of Ophthalmology\\
University of Illinois Chicago\\
Chicago, IL, USA \\
\And
Ann Q. Tran \\
Department of Ophthalmology\\
University of Illinois Chicago\\
Chicago, IL, USA \\
}

\begin{document}
\maketitle
\begin{abstract}
Periorbital measurements such as margin reflex distances (MRD1/2), palpebral fissure height, and scleral show are critical in diagnosing and managing conditions like ptosis and disorders of the eyelid. We developed and evaluated Glorbit, a lightweight, browser-based application for automated periorbital distance measurement using artificial intelligence (AI), designed for deployment in low-resource clinical environments. The goal was to assess its usability, cross-platform functionality, and readiness for real-world field deployment. The application integrates a DeepLabV3 segmentation model into a modular image processing pipeline with secure, site-specific Google Cloud storage. Glorbit supports offline mode, local preprocessing, and cloud upload through Firebase-authenticated logins. The full workflow—metadata entry, facial image capture, segmentation, and upload—was tested. Post-session, participants completed a Likert-style usability survey. Glorbit successfully ran on all tested platforms, including laptops, tablets, and mobile phones across major browsers. A total of 15 volunteers were enrolled in this study where the app completed the full workflow without error on 100\% of patients. The segmentation model succeeded on all images, and average session duration was 101.7 ± 17.5 seconds. Usability scores on a 5-point Likert scale were uniformly high: intuitiveness and efficiency (5.0 ± 0.0), workflow clarity (4.8 ± 0.4), output confidence (4.9 ± 0.3), and clinical usability (4.9 ± 0.3). mGlorbit is a functional, cross-platform solution for standardized periorbital measurement in clinical and low-resource settings. By combining a local image processing with secure, modular data storage and offline compatibility, the tool enables scalable deployment and secure data collection. These features support broader efforts in AI-driven oculoplastics including future development of real-time triage tools and multimodal datasets for personalized ophthalmic care. The tool can be accessed at \url{https://glorbit.app}.
\end{abstract}

\section{Introduction}
Periorbital measurements such as margin to reflex distance 1 and 2 (MRD 1/2), palpebral fissure height, and scleral show are essential components of clinical assessment in a range of conditions including ptosis, thyroid eye disease, and congenital craniofacial conditions.\cite{conrady_crouzon_2025,koka_ptosis_2025,maclachlan_normal_2002} These measurements guide surgical decision-making, track disease progression, and serve as critical inputs for diagnosis.\cite{nemet_accuracy_2015,cruz_digital_1998,cruz_quantification_1999,bodnar_automated_2016} However, in many settings, particularly in rural or resource-limited environments, such measurements are performed inconsistently or not at all due to lack of trained personnel or standardized tools.8,9 In the US alone, 89\% of counties do not have a single oculoplastic surgeon, the primary physician responsible for obtaining such measurements clinically.\cite{hussey_oculofacial_2022}

Previous work by our group demonstrated that automated periorbital measurements from facial photographs can achieve clinically acceptable accuracy using deep learning-based segmentation approaches.\cite{nahass_open-source_2025,nahass_state---art_2024} A trained DeepLabV3 model achieved mean absolute errors consistently below established intergrader variability thresholds for key measurements including MRD1, MRD2, and intercanthal distance, with 86\% of measurements falling within intergrader thresholds. Importantly, this approach demonstrated superior robustness compared to the previous state of the art method, periorbitAI, successfully processing 100\% of images across diverse disease populations including thyroid eye disease and craniofacial conditions, compared to 59-85\% success rates.\cite{nahass_state---art_2024,van_brummen_periorbitai_2021} However, these algorithms remained confined to research settings due to the lack of deployable infrastructure capable of integrating image capture, processing, and secure data management into clinical workflows.

In response to this gap, we developed Glorbit, a web-based application for AI based periorbital distance measurement and metadata capture. The platform is designed for deployment in clinical environments with limited infrastructure and supports secure, site-specific cloud storage through Google Cloud integration. In this paper, we describe the system architecture and image processing pipeline and demonstrate the functional readiness of the app for field use in both academic and global health settings.

\section{Methods}
\subsection{System Overview and User Flow}
\begin{figure}
    \centering
    \includegraphics[width=1\linewidth]{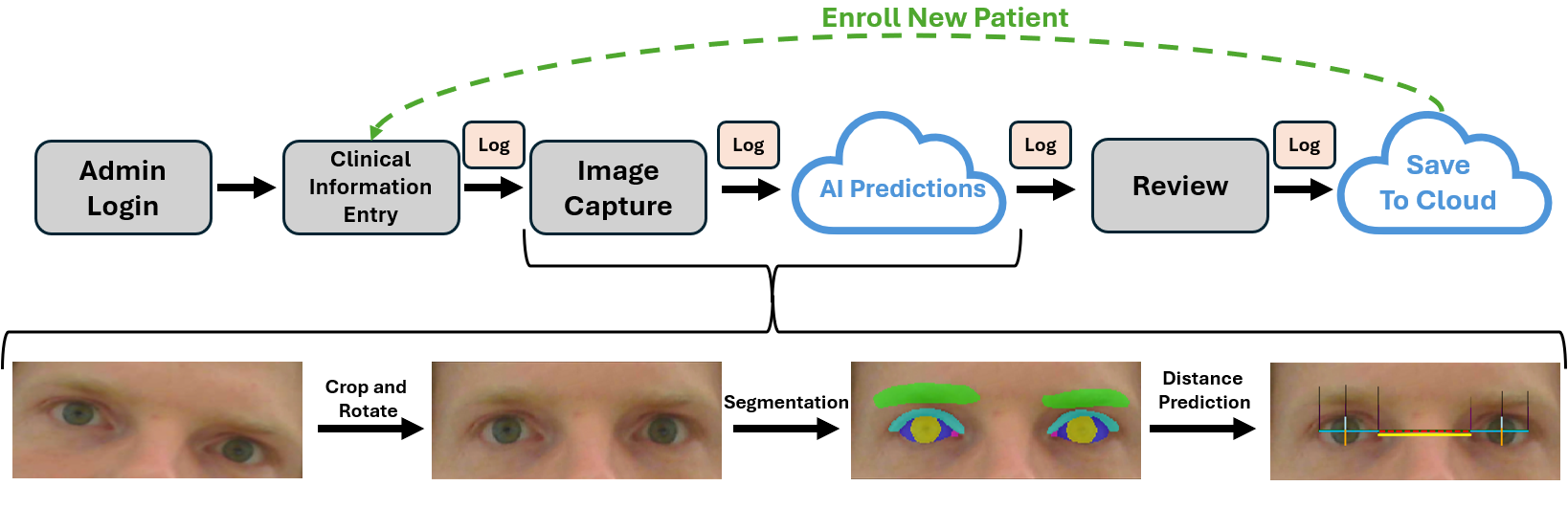}
    \caption{Glorbit application workflow and AI processing pipeline. The top row illustrates the user-facing app interface: a user logs in, enters clinical metadata, captures an image, reviews AI-predicted periorbital distances, and uploads results to cloud storage. Cloud-shaped nodes denote steps involving remote model inference or cloud-based data storage. All user actions, including reverse navigation, are logged. The bottom row depicts the backend image processing steps: cropping and rotation based on facial landmarks, semantic segmentation of key anatomical regions, and automated distance prediction using landmark geometry.}
    \label{fig:fig_1}
\end{figure}

Glorbit is a browser-based application designed to capture standardized periorbital measurements in clinical and low-resource environments. Upon login, authenticated users are presented with a structured form to collect brief patient history and visit metadata. This information is stored temporarily in session state within the browser. After submission, users proceed to an image capture page, where a webcam or device camera is used to capture a frontal facial image. The image is processed locally to generate segmentation and distance predictions, which are then displayed for user review. Once confirmed, the image and associated metadata are uploaded to a secure, site-specific cloud storage location. Logging is integrated throughout the pipeline to capture system events, errors, and upload status. See Figure \ref{fig:fig_1} for an overview of users steps throughout the application.

\subsection{Image Processing Pipeline}
Each captured image undergoes a standardized pipeline for anatomical alignment, cropping, and segmentation described by Nahass et. al.\cite{nahass_state---art_2024} MediaPipe FaceMesh landmarks are used to rotate and align the image to a horizontal eye-level axis.\cite{kartynnik_real-time_2019} A region of interest is then cropped around the eyes and periorbital region. The aligned image is passed to a trained DeepLabV3 segmentation model for periorbital anatomic segmentation, followed by measurement of key periorbital distances from the segmented output (Figure \ref{fig:fig_1}).\cite{nahass_state---art_2024} In the event of a model failure, the app handles the error, logs the failure, and prompts the user to try again.

\subsection{Clinical Data}
Glorbit captures a structured set of clinical metadata alongside each image to support contextual analysis and downstream modeling. At minimum, users are required to input a deidentified patient ID, age, and sex to proceed with the workflow. Optional fields include ethnicity (selected from a site-customizable dropdown list), relevant comorbidities (e.g., thyroid eye disease, craniofacial conditions, prior eyelid surgery), visit type (new or follow-up), and image conditions (e.g., use of tape, lighting notes). Sites may also enable entry of laboratory values when available. Supported labs include ALT, AST, HbA1c, eGFR, TSH, TSI, ACR, and calcium, with standardized units displayed next to each input. All metadata is stored alongside the captured image and predicted periorbital measurements in the designated site-specific cloud storage bucket. Only the required fields must be completed to advance to image capture; all other fields are optional and may be tailored to institutional needs.

\subsection{Handling and Storage Architecture}
\begin{figure}
    \centering
    \includegraphics[width=1\linewidth]{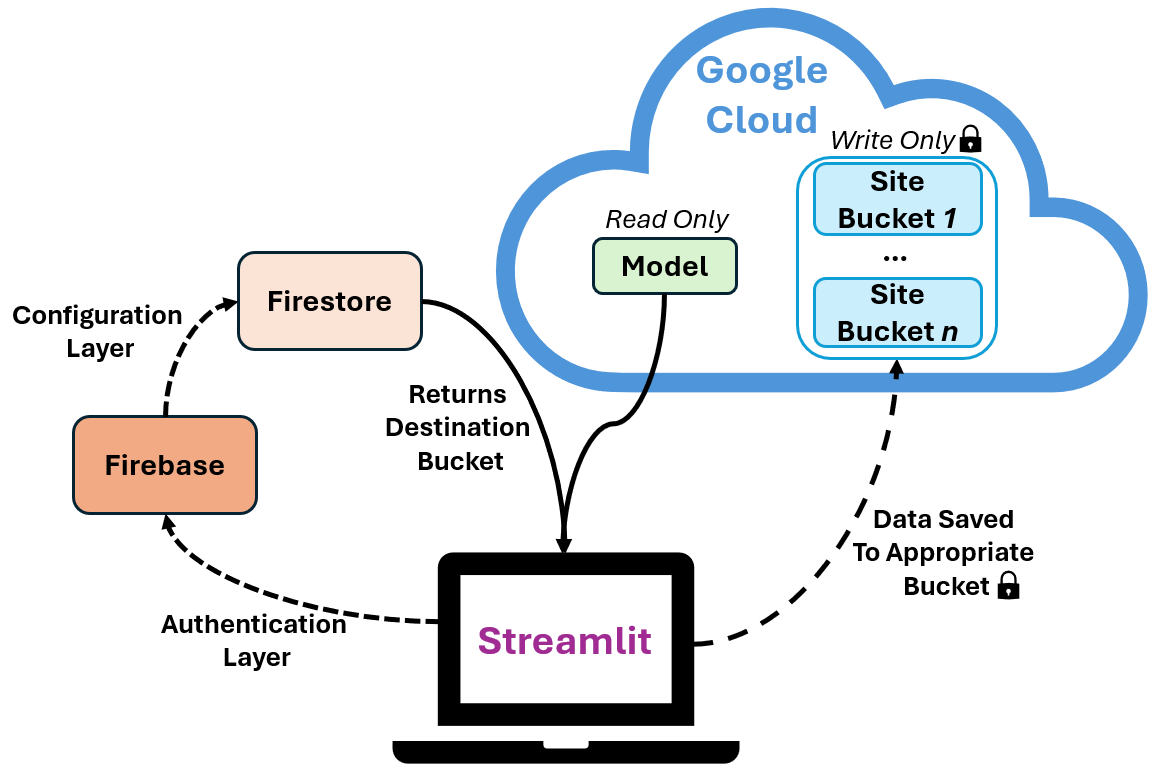}
    \caption{System architecture for secure authentication, configuration, and cloud-based storage within the Glorbit platform. Users authenticate via Firebase, after which Firestore provides site-specific configuration, including the destination storage bucket. The local Streamlit app reads AI model weights from a centralized, read-only location in Google Cloud and writes collected data to site-specific, write-only buckets. All data transfers are protected using encryption at rest and in transit, as indicated by lock icons. Access control is governed via Google Cloud’s Identity and Access Management (IAM) framework. For offline use, the system can operate without Firebase or Firestore by manually specifying user credentials and destination storage paths.}
    \label{fig:fig_2}
\end{figure}

The app frontend is built in Streamlit and deployed as a Docker container to ensure reproducibility and offline compatibility. Firebase Authentication manages secure login, with role-based access control and per-site credentialing.  Firestore is used to assign each user’s upload path based on their authenticated credentials, enabling dynamic configuration of storage destinations. This allows a single instance of the app to be deployed across multiple institutions without requiring hardcoded site-specific settings or local reconfiguration. Uploaded data, including images, predicted measurements, logs, and metadata, is stored in Google Cloud Storage (GCS), with each site allocated a dedicated storage bucket. All data is encrypted at rest and in transit, and access rules are configured through Google Cloud IAM policies. Glorbit is optimized for deployment on consumer-grade hardware and can be accessed on laptops, tablets, and mobile phones with integrated webcams. In settings where internet is not available, the application can be run offline with all data stored locally. In offline mode, the application is shipped in a docker container with all model weights included locally. A graphical schematic of data movement from the user’s perspective can be found in Figure \ref{fig:fig_2}.

\subsection{Security Design}
No personally identifiable information is collected beyond a user-entered deidentified patient ID and cropped images of the eyes. All uploads are encrypted both in transit and at rest using Google Cloud’s native encryption. Each deployment site is assigned a unique Google Cloud Storage bucket with IAM roles that can be configured to restrict access by site. Firebase Authentication enforces user-specific logins, and Firestore assigns site-level write access dynamically. Full event and error logging is implemented for auditability. While Glorbit is designed to minimize the handling of protected health information (PHI), the infrastructure is compatible with HIPAA requirements and, when deployed under an institutional Business Associate Agreement (BAA), constitutes a fully HIPAA-secure system.\cite{noauthor_hipaa_nodate}

\subsection{Internal Testing: Stress Testing and Cross-Platform Compatibility}
In addition to the simulated enrollment study, the app was manually tested under expected failure conditions by a member of the study team (GN). These included poor lighting, misaligned or undetectable faces, segmentation model failures, and network interruptions during upload. The goal was to verify that the app responded appropriately to common sources of real-world failure. associated events for auditability. To confirm cross-platform compatibility, the app was also run across major browsers (Chrome, Firefox, Safari) and device types (Windows laptops, tablets, Chromebooks, and smartphones). 

\subsection{External Testing: Survey and Simulated Enrollment Testing}
To evaluate usability, we conducted a simulated enrollment study with 15 adult volunteers (age $\geq$ 18) at the Illinois Eye and Ear Infirmary. No identifying information was collected, and each participant was assigned a deidentified patient ID. Verbal consent was obtained in accordance with an IRB-exempt protocol (STUDY2025-0731).

Each session was conducted by a trained study operator (GRN) using a 2021 MacBook Air in a private room under natural indoor lighting conditions. The operator followed the standard Glorbit workflow, including metadata entry and facial image capture. All data, including images and metadata, were permanently deleted immediately after each session.

Following the simulated interaction, participants completed a brief anonymous survey assessing usability. The survey included five Likert-scale items (1–5) evaluating intuitiveness, efficiency, clarity of workflow, confidence in outputs, and perceived clinical utility. All responses were stored without any identifiers.

\section{Results}
\subsection{App Walkthrough and Failure Mode Evaluation}
\begin{figure}
    \centering
    \includegraphics[width=1\linewidth]{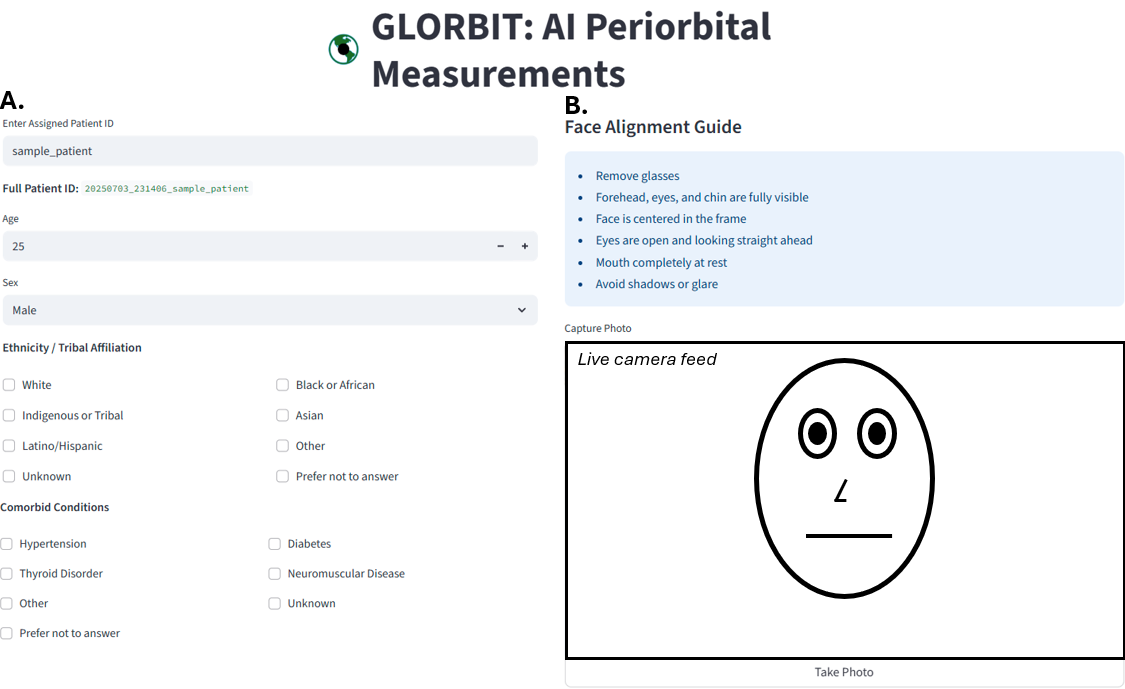}
    \caption{Screenshot of the Glorbit app interface. (A) The form used to input patient metadata, including demographics and comorbidities. (B) Face alignment guide with a live camera view for standardized image capture. These two steps represent the only direct patient interactions aside from the final review and submission step.}
    \label{fig:fig_3}
\end{figure}

Screenshots of the core workflow of Glorbit are shown in Figure \ref{fig:fig_3}, illustrating the minimalistic, stepwise interface designed for ease of use in clinical environments.

To ensure robustness, we manually tested the application under expected failure scenarios, including poor lighting, undetectable faces, segmentation errors, and interrupted uploads. In each case, the system produced appropriate user-facing messages, logged the event, and preserved metadata.

\subsection{Cross Platform Compatibility}
The web-based application demonstrated consistent functionality across major browsers (Chrome, Firefox, Safari) and device types including Windows laptops, mobile devices, tablets, and Chromebooks. All core features- authentication, form entry, camera access, and image processing- operated without modification across platforms.

\subsection{Simulated Enrollment Performance}
We enrolled 15 adult volunteers in a simulated usage test to assess the practical performance of Glorbit under realistic operating conditions. Participant demographics are summarized in Table \ref{tab:metadata}. The group’s mean age was 32.6 ± 11.0 years, was 67\% female, and spanned five self-identified ethnic groups.

 In all 15 cases (100\%), the app completed the full workflow from metadata entry, image capture, segmentation, to upload without crashing. The segmentation model returned successful results in all cases, and no image processing or upload failures were observed. The average session duration, measured from initial form entry to final upload, was 101.7 ± 17.5 seconds on consumer-grade hardware. Logging was successful in all cases, with system events and session metadata stored as expected.

\begin{table}[]
\centering
\resizebox{.5\columnwidth}{!}{%
\begin{tabular}{@{}|c|cc|@{}}
\toprule
\cellcolor[HTML]{FFFFFF}\textbf{}    & \multicolumn{2}{c|}{\textbf{n=15}}                  \\ \midrule
                                     & \multicolumn{1}{c|}{Ave ± Std}        & 32.6 ± 11.0 \\ \cmidrule(l){2-3} 
                                     & \multicolumn{1}{c|}{Max}              & 58          \\ \cmidrule(l){2-3} 
                                     & \multicolumn{1}{c|}{Min}              & 24          \\ \cmidrule(l){2-3} 
\multirow{-4}{*}{\textbf{Age}}       & \multicolumn{1}{c|}{Med}              & 26          \\ \midrule
                                     & \multicolumn{1}{c|}{Male}             & 33\%        \\ \cmidrule(l){2-3} 
\multirow{-2}{*}{\textbf{Sex}}       & \multicolumn{1}{c|}{Female}           & 67\%        \\ \midrule
                                     & \multicolumn{1}{c|}{White}            & 13\%        \\ \cmidrule(l){2-3} 
                                     & \multicolumn{1}{c|}{Latino/Hispanic}  & 27\%        \\ \cmidrule(l){2-3} 
                                     & \multicolumn{1}{c|}{Black or African} & 20\%        \\ \cmidrule(l){2-3} 
                                     & \multicolumn{1}{c|}{Asian}            & 33\%        \\ \cmidrule(l){2-3} 
\multirow{-5}{*}{\textbf{Ethnicity}} & \multicolumn{1}{c|}{Other}            & 7\%         \\ \bottomrule
\end{tabular}%
}
\caption{Demographic characteristics of simulated study participants (n = 15). Age is reported as mean ± standard deviation (SD), with additional summary statistics. Sex and ethnicity are reported as percentages of the total participant group.}
\label{tab:metadata}
\end{table}

\subsection{User Feedback}
Following each session, participants completed a brief anonymous post-usage survey to assess their experience using the app. Average responses were uniformly high across all five items (Table 2\ref{tab:feedback}). Participants rated the app as intuitive and efficient (both 5.0 ± 0.0), with strong confidence in the displayed outputs (4.9 ± 0.3) and a high level of comfort using the tool in a clinical setting (4.9 ± 0.3). Participants also reported clear understanding of each step in the data collection workflow (4.8 ± 0.4).

\begin{table}[!ht]
\centering
\resizebox{\columnwidth}{!}{%
\begin{tabular}{@{}|c|c|@{}}
\toprule
\cellcolor[HTML]{FFFFFF}\textbf{Statement}               & Average Rating (mean ±   SD) \\ \midrule
The Glorbit app was   intuitive and easy to use.         & 5.0 ± 0.0                    \\ \midrule
The process of capturing   and uploading patient data was fast and efficient. & 5.0 ± 0.0 \\ \midrule
I would feel comfortable   using this tool in a clinical setting.             & 4.9 ± 0.3 \\ \midrule
I understood each step   of the data collection process. & 4.8 ± 0.4                    \\ \midrule
I have confidence in the   displayed outputs.            & 4.9 ± 0.3                    \\ \bottomrule
\end{tabular}%
}
\caption{Summary of user feedback from post-session surveys in the simulated enrollment study. Each participant (n = 15) rated five statements on a 5-point Likert scale (1 = strongly disagree, 5 = strongly agree). Responses reflect perceived usability, efficiency, clarity, confidence, and clinical applicability of the Glorbit app. Values are reported as mean ± standard deviation.}
\label{tab:feedback}
\end{table}

\section{Discussion}
Many periorbital distance prediction algorithms have been developed in recent years, but their clinical impact remains limited by the absence of deployable infrastructure, particularly in low-resource settings.\cite{hussey_oculofacial_2022,van_brummen_periorbitai_2021,rana_artificial_2024,chen_smartphone-based_2021,guo_novel_2021} In such contexts, successful AI deployment demands lightweight, secure, and field-ready systems capable of image capture, processing, and storage. This need is especially pressing in subspecialties such as oculoplastic and craniofacial surgery, where access to trained providers is scarce.\cite{hussey_oculofacial_2022}

To meet this need, we developed an internet browser-based app, Glorbit. In this feasibility study of 15 participants, Glorbit was successful in metadata entry, image capture, segmentation, and upload. It also performed well across different internet browsers and device types. Importantly, Glorbit provides error messages and logs all user movement in the event of a failure (which was not observed in the simulated user study). This comprehensive error handling ensures system reliability and provides administrators with detailed diagnostic information for troubleshooting deployment issues.

To date, several mobile applications for facial measurement have been developed, but these lack the comprehensive features required for clinical deployment are mainly focused on interpupillary distance for use in optometry clinics and glasses fitting or generic facial landmark detection.\cite{han_comparing_nodate,noauthor_bonlook_nodate,noauthor_eyeque_2024,singman_accuracy_2014} Existing solutions typically offer only basic measurement capabilities without secure multi-site data management, offline functionality, or integration with cloud storage infrastructure.\cite{han_comparing_nodate,singman_accuracy_2014} Most importantly, no current platform combines automated AI-based measurement with the flexible deployment architecture needed for diverse clinical environments. A comparison table of research and consumer products with Glorbit can be found in Table \ref{tab:comp}. Glorbit addresses these limitations through a combination of offline/online capability, site-specific secure cloud storage, and modular architecture that can be deployed across institutions without requiring local IT infrastructure.

\begin{table}[]
\centering
\resizebox{\columnwidth}{!}{%
\begin{tabular}{@{}|c|c|c|c|c|c|@{}}
\toprule
\cellcolor[HTML]{FFFFFF}\textbf{Platform} & \textbf{Offline Support} & \textbf{Cloud Storage} & \textbf{For Clinical Use} & \textbf{Measurement Type} & \textbf{Notes} \\ \midrule
\textbf{Glorbit}       & Yes & Yes & Yes & MRD1, MRD2, PFH, (48 distances) & Site-specific secure storage,   open-source, real-time feedback        \\ \midrule
\textbf{MediaPipe\cite{kartynnik_real-time_2019}}   & Yes & No  & No  & Facial landmarks                & Google tool for face mesh   detection                                  \\ \midrule
\textbf{PDCheck AR\cite{noauthor_eyeque_2024}}  & No  & No  & Yes & IPD                             & Mobile app for interpupillary   distance (IPD) measurement             \\ \midrule
\textbf{GlassesOn\cite{noauthor_glasseson_nodate}}   & Yes & No  & Yes & IPD                             & Consumer app for virtual try-on   and IPD estimation                   \\ \midrule
\textbf{OpenFace\cite{baltrusaitis_openface_2016}} & Yes & No  & No  & Facial landmarks                & Academic tool for facial   behavior analysis, can extract 68 landmarks \\ \midrule
\textbf{Face++\cite{noauthor_face_nodate}}      & No  & Yes & No  & Facial geometry                 & Commercial API for face   analysis, includes eye/brow landmarks        \\ \bottomrule
\end{tabular}%
}
\caption{Comparison of Glorbit to existing facial measurement platforms. "Offline Support" indicates whether the platform can run without internet access. "Cloud Storage" reflects availability of integrated cloud upload functionality. "For Clinical Use" refers to whether the platform is designed for clinical environments or use cases. "Measurement Type" describes the anatomical features or distances quantified by each system. Glorbit uniquely supports automated periorbital distance prediction across 48 anatomical features, secure cloud integration, offline operation, and deployment in clinical workflows. }
\label{tab:comp}
\end{table}

In addition to serving as a storage repository for providers, the captured data can also be used as future training data to create patient-specific models. One of the most persistent barriers to equitable clinical AI in remote settings is poor model generalization to underrepresented populations, often due to data captured under different conditions being out of distribution to the original training set.\cite{hong_out--distribution_2024,rashidisabet_validating_2023,martensson_reliability_2020,ozkan_multi-domain_2024,karimi_improving_2023} Glorbit addresses this challenge by enabling ethically governed data collection across diverse global sites. By embedding image capture and metadata collection into clinical workflows, the platform supports the development of more representative, generalizable models through real-world data acquisition and iteration. Further strengthening the motivation for large scale collection of high quality data is the developing field of Oculomics which aims to leverage deep learning to make predictions about systemic health from images of the eyes.\cite{zhu_oculomics_2025,honavar_oculomics_2022,patterson_oculomics_2024,suh_oculomics_2024,zhou_foundation_2023} While most current Oculomics research focuses on retinal imaging, recent studies have shown that external eye photographs can also be used to predict systemic biomarkers such as ALT, AST, and HBA1C.\cite{babenko_deep_2023,debuc_ai_2023,babenko_detection_2022} With the increasing number of foundational ophthalmic deep learning systems, having specific multimodal datasets available for finetuning may enable the development of more precise clinical AI tools.

An ongoing challenge of integrating tools like Glorbit into clinical workflows is protection of patient privacy and compliance with institutional and national data governance requirements. This includes obtaining ethical approval for clinical use and, where applicable, establishing a Business Associate Agreement (BAA) between the deploying institution and the cloud provider. When operated under such a BAA, Glorbit’s architecture, built on HIPAA-compliant infrastructure with encryption, access control, and audit logging, constitutes a fully HIPAA-compliant system.\cite{noauthor_hipaa_nodate,isola_protected_2025} While the app itself collects minimal PHI, institutional oversight remains a critical component of medical AI for the foreseeable future. To promote transparency and allow for local adaptation and compliance with patient protection laws on a per nation basis, Glorbit is fully open-source. Sites can deploy the app locally without cloud access or configure their own cloud backend with minimal engineering effort. 

While Glorbit is currently designed as a research and data collection platform, the modular architecture enables expansion of clinical utility through additional features such as clinician-facing dashboards for monitoring periorbital measurements over time, as well as the integration of lightweight triage models capable of flagging eyes with features suggestive of disease. Furthermore, our simulated enrollment study demonstrated that Glorbit is both usable and efficient in real-world clinical conditions. Participants consistently rated the interface as intuitive and the image capture process as quick, with high levels of confidence in the predicted measurements. The average time to complete a full session was under 2 minutes, reinforcing its feasibility for routine clinical use. These findings support the platform’s potential as a scalable solution for integrating AI-assisted measurement into resource-constrained environments. Prior studies have shown that craniofacial and oculoplastic conditions can be accurately classified from periorbital distances.\cite{nahass_state---art_2024} As Glorbit evolves, such models can be embedded directly into the interface to provide real-time clinical decision support and triage recommendations. Future work includes field deployment of Glorbit in diverse global health environments, such as rural hospitals, to evaluate performance in real-world clinical workflows

\subsection{Acknowledgements}
The study team would like to gratefully acknowledge Alexander Schönjahn for deployment assistance at Quina Care. We also thank our funders. This study was supported by an unrestricted grant from Research to Prevent Blindness, a donation from the Cless Family Foundation, and the National Institutes of Health P30 EY001792 core grant. The funding organizations had no role in the design or conduct of the study; collection, management, analysis, or interpretation of the data; preparation, review, or approval of the manuscript; or decision to submit the manuscript for publication.

\subsection{Conflict of Interest Disclosures:}

Dr. Hubschman reports stock or stock options in Horizon Surgical Systems. Dr. Setabutr reports consultancy for Oyster Point Pharma and is a board member of the American Society of Ophthalmic Plastic and Reconstructive Surgery; he also reports stock or stock options in Lodestone Pharmaceuticals. Dr. Tran reports grants from the ISPB and consultancy for Genentech/Roche. No other disclosures were reported.

\subsection{Abbreviations}

AI: artificial intelligence; ALT: alanine aminotransferase; AST: aspartate aminotransferase; BAA: Business Associate Agreement; eGFR: estimated glomerular filtration rate; GCS: Google Cloud Storage; HIPAA: Health Insurance Portability and Accountability Act; IAM: Identity and Access Management; IRB: Institutional Review Board; MRD1/2: margin reflex distance 1 and 2; PFH: palpebral fissure height; PHI: protected health information; SD: standard deviation; TSH: thyroid-stimulating hormone; TSI: thyroid-stimulating immunoglobulin; HbA1c: hemoglobin A1c.

\bibliographystyle{unsrt}
\bibliography{glorbit}

% \bibliographystyle{unsrt}  
% %\bibliography{references}  %%% Remove comment to use the external .bib file (using bibtex).
% %%% and comment out the ``thebibliography'' section.

% %%% Comment out this section when you \bibliography{references} is enabled.
% \begin{thebibliography}{1}

% \bibitem{kour2014real}
% George Kour and Raid Saabne.
% \newblock Real-time segmentation of on-line handwritten arabic script.
% \newblock In {\em Frontiers in Handwriting Recognition (ICFHR), 2014 14th
%   International Conference on}, pages 417--422. IEEE, 2014.

% \bibitem{kour2014fast}
% George Kour and Raid Saabne.
% \newblock Fast classification of handwritten on-line arabic characters.
% \newblock In {\em Soft Computing and Pattern Recognition (SoCPaR), 2014 6th
%   International Conference of}, pages 312--318. IEEE, 2014.

% \bibitem{hadash2018estimate}
% Guy Hadash, Einat Kermany, Boaz Carmeli, Ofer Lavi, George Kour, and Alon
%   Jacovi.
% \newblock Estimate and replace: A novel approach to integrating deep neural
%   networks with existing applications.
% \newblock {\em arXiv preprint arXiv:1804.09028}, 2018.

% \end{thebibliography}

\end{document}